\documentclass[a4paper,12pt]{article}
\usepackage{amsmath,amssymb,amsfonts}
\usepackage{authblk}
\usepackage{graphicx}
\usepackage{geometry}
\title{A method for obtaining nonspreading solutions of the Schr\"{o}dinger equation}
\author[1]{Tamoghna Majumdar \thanks{tamz.msc1@gmail.com}}
\author[2]{Maitraya Kanta Bhattacharyya \thanks{mkb14ip002@iiserkol.ac.in}}
\author[1,2]{K. Rajesh Nayak \thanks{rajesh@iiserkol.ac.in}}
\affil[1]{\normalsize \textit{Center of Excellence in Space Sciences India, Indian Institute of Science Education and Research, Kolkata}}
\affil[2]{\textit{Department of Physical Sciences, IISER Kolkata, Mohanpur: 741246, India}}
\everymath{\displaystyle}
\begin{document}

\maketitle

\begin{abstract}
 In this paper, we present a simple analytical method for obtaining a nonspreading solution of the time-dependent Schr\"{o}dinger equation, which is given by the Airy function. 
 The solution is derived by imposing a restriction on the phase factor of the ansatz that is taken to solve the differential equation. 
 Considering at first the free particle, we show that nonspreading solutions can also be obtained for a time-dependent linear potential. 
 The method is shown to work in both one and two dimensions, and can be  easily extended if required. The applicability of the method is discussed in 
 relation to the nonlinear case. 
\end{abstract}

\section{Introduction}
Berry and Balazs showed   that the free-particle Schr\"{o}dinger equation admits non-trivial solutions whose amplitude is given by the Airy differential equation~\cite{berry}. 
Specifically, they showed that 
\begin{equation} 
 -\frac{\hbar^2}{2m}\psi_{xx} = i\hbar\psi_{t} \,,
 \label{eq:1}
\end{equation}
has a unique solution given by
\begin{equation}
 \psi(x,t) = \mathrm{Ai}\left[\frac{B}{\hbar^{2/3}}\left(x-\frac{B^3t^2}{4m^2}\right)\right]e^{\left(iB^3t/2m\hbar\right)\left[x-\left(B^3t^2/6m^2\right)\right]}\,,
 \label{eq:2}
\end{equation}
where $\mathrm{Ai}(z)$ is the Airy function and $B$ is a positive constant. Unlike plane wave solution, the  Airy  solution  given above is not separable in the variables $\{x,\, t\}$. The essential feature of these solutions is that the time-evolution of the wavepacket is non-dispersive.
The solution given by Eq.~(\ref{eq:2}) was proved to be the only unique solution of Eq.~(\ref{eq:1}) having this property.
This result was originally explained in terms of the behaviour of the corresponding families of the semiclassical orbits in phase space. Subsequently, much work has been done
to interpret this surprising result, with a proper quantum mechanical derivation being given in~\cite{unni}. 

Recently, the nonspreading solution has evoked considerable interest in
the community. It has been shown that, under the paraxial approximation
the wave equation allows a nonspreading solution~\cite{ref:Siviloglou1}. It has been verified experimentally by generating optical beams with wave packets that are accelerated~\cite{ref:Siviloglou}. This leads further possibility of using these solutions in various optical systems. 

In this paper,  we show that it is possible to obtain the result from techniques
based on ordinary calculus. It is shown that the in the ansatz $A(x,t)e^{i\phi(x,t)}$ that is used to obtain wave-like solutions of Eq.~(\ref{eq:1}), a particular choice of
the phase $\phi$ will lead to a reduction of the Schr\"{o}dinger equation to the Airy differential equation.
In this paper, we first describe a method for obtaining the nonspreading solution for  the Schr\"{o}dinger equation for a  free particle in one dimension. We also show that we can easily extend the solution for linear and time-dependent potentials.  In section~\ref{sec:3}, we show that the method can be extended to higher dimension by taking the two dimensional case as an example. 
\section{The method for obtaining solution\label{sec:2}}
We are looking for non-trivial solutions of 
\begin{equation}
 \psi_{xx} + \mathrm{i}\kappa\psi_t - \bar{V}(x,t)\psi = 0\,,
 \label{eq:sceqgen}
\end{equation}
where $\kappa = 2m/\hbar$ and $\bar{V} = (\kappa/\hbar)V$. We start with the free particle case first with $\bar{V}(x,t)=0$. After demonstrating the method, we extend the solution  with a potential term. 
\subsection{The one-dimensional Schr\"{o}dinger equation}
In this section we solve for the non-disperssive  solution in one dimension. We start with   $V=0$ in Eq.~(\ref{eq:sceqgen}). First, we propose a trial solution of the form,  
\begin{equation}
\psi = A(x,t)\,e^{i\phi(x,t)}\,,
\label{eq:modelson}
\end{equation}
 where $A(x,t)$ and $\phi(x,t)$ are assumed to be real functions. Separating real and imaginary parts after substituting Eq.~(\ref{eq:modelson}) in Eq.~(\ref{eq:sceqgen}), we obtain:
 \begin{eqnarray}
  A_{xx} - A(\phi_x)^2 -\kappa A \phi_t & = & 0\,, \nonumber \\
  A\phi_{xx} + 2 A_x\phi_x + \kappa A_t &= &0\,.
  \label{eq:Aphi}
\end{eqnarray}
We start with the {\it ansatz} that the phase term $\phi(x,t)$ must satisfy the 
Laplace equation,
\begin{equation}
 \frac{\partial^2\phi}{\partial x^2} = 0\,, \label{eq:lap=0}
\end{equation}
With this  we get,  
\begin{equation}
\phi = \phi_1 \cdot x + \phi_0\,,
\label{eq:7}
\end{equation}
 where $\{\phi_1,\phi_0\}$ are functions of time only. 
Substituting $\phi$ in Eq.~(\ref{eq:Aphi}), we get
\begin{equation}
 A_{xx} - Ay = 0\,,
\end{equation}
where, 
\begin{equation}
y = \phi_1^2+\kappa\left(\dot{\phi}_1 x+\dot{\phi}_0\right)\,.
\end{equation}
 In the above, $\dot{\phi}=\frac{d\phi}{dt}$. We further assume that $\partial y/\partial x$ is constant implying $\dot{\phi}_1 = P$ (a constant). Then we have reduced Eq.~(\ref{eq:sceqgen}) to an Airy differential equation with the solution
\begin{equation}
 A = \mathrm{Ai}\left[\frac{y}{(\kappa P)^{2/3}}\right]\,.
\end{equation}
The condition for $\phi_0$ is given from Eq.~(\ref{eq:Aphi}) by
\begin{equation}
 \kappa \ddot{\phi}_0 + 4P^2t  = 0\,, 
\end{equation}
resulting in,
\begin{equation}
 \phi_0 = -\frac{2}{3\kappa}P^2t^3 + Qt\,.
\end{equation}
Here $P$ and $Q$ are constants. It can be shown that one can get Berry's solution with $Q=0$ and $P=B^3/2m\hbar$ where $B$ is an arbitrary positive
 constant.

At this point, we would like to physically justify the {\it ansatz} used in obtaining the solution. It is  closely related to  the symmetry properties of the Schr\"{o}dinger equation. The Schr\"{o}dinger equation admits a 
Galilean transformation which preserves the form of the differential equation \cite{greenberger2}. The transformed wave-function is then just the original wave-function multiplied  by a phase which is linear in the spatial coordinates. Thus, we must have $\partial^2 \phi/ \partial x^2 = 0$.

\subsection{With linear potential\label{sec:linpot}}
In this section,  we  generalise to the case with potential. Starting with the same trial solution to obtain, 
\begin{eqnarray}
  A_{xx} - A(\phi_x)^2 -\kappa A \phi_t - \bar{V}A = 0\,, \nonumber \\
  A\phi_{xx} + 2 A_x \phi_x + \kappa A_t = 0\,.
\end{eqnarray}
Since the second equation is the same as given in Eq.~(\ref{eq:Aphi}),  the form of $\phi$ is modified by  an additional term due to the potential. The equation for the amplitude $A$ also takes the  same form in terms of the variable $y = \phi_{1}^2+\kappa\left(\dot{\phi}_1 x + \dot{\phi}_0\right) + \bar{V}$,  as before, $\partial y/\partial x$ is assumed constant. Like before, 
we take $\mathrm{d}\phi_{1}/\mathrm{d}t = P$ with the addition that $\partial\bar{V}/\partial x = \bar{V_1}$, both of them being constants. With these conditions, we have
\begin{equation} 
\frac{\partial y}{\partial x} = \kappa \frac{\mathrm{d}\phi_{1}}{\mathrm{d}t} + \frac{\partial\bar{V}}{\partial x} = \kappa P + \bar{V_1}\,, 
\end{equation}
from which we finally get 
\begin{equation}
\phi_1 = Pt~,\qquad \bar{V} = \bar{V_1}x + \bar{V_0}\,.
\end{equation}
Here, we have freedom to choose $\bar{V_0}$ to be a function of time. In order to get $\phi_0$ we solve an equation similar to Eq.~(\ref{eq:7}) to obtain
\begin{equation}
 \phi_0 = -\frac{1}{\kappa^2}\left[\int^t\bar{V_0}(t')\mathrm{d}t' + \frac{2}{3}\kappa P^2t^3 -\frac{1}{3}P\bar{V_1}t^3\right] + Qt~,
\end{equation}
where $Q$ is a constant. It can be readily seen this is a validation of the results of Berry and Balazs for more general potentials of the form $F(t)x$.
\section{Solution  in two dimensions\label{sec:3}}
In this section, we  consider the Schr\"{o}dinger equation in two spatial dimensions. For the potential-free case we wish to solve,
\begin{equation}
\psi_{xx} + \psi_{yy} + i\kappa\psi_t = 0\,, 
\end{equation}
with the trial solution $\psi = A(x,y,t)e^{i\psi(x,y,t)}.$ Proceeding similarly as before, the real and imaginary parts are separated as
\begin{eqnarray}
 A_{xx} + A_{yy} - A(\phi_x)^2 - A(\phi_y)^2 - \kappa A \phi_t &=& 0\,, 
 \nonumber\\
 A\phi_{xx} + A\phi_{yy} + 2A_x\phi_x + 2A_y\phi_y + \kappa A_t & = &0\,.
 \label{eq:2dsep}
\end{eqnarray}
Thus,  the required conditions on $\phi$ are $\phi_{xx}=\phi_{yy}=0$, which gives us 
\begin{equation}
\phi = \phi_0 + \phi_1 x + \phi_2 y\,.
\end{equation}
Here all $\phi_i$ $(i=0,1,2)$ are functions of time only. Substituting back into Eq. (\ref{eq:2dsep}) we get
\begin{equation} 
A_{xx} + A_{yy} - Az = 0\,,
\end{equation}
where, 
\begin{equation}
z = \phi_{1}^2 + \phi_{2}^2  + \kappa \left(\dot{\phi}_1 x + \dot{\phi}_2 y +\dot{\phi}_0 \right)\,.
\end{equation}
Thus, the required solution for the amplitude is 
\begin{equation}
A = \mathrm{Ai}\left[\frac{z}{\kappa^{2/3}(P_{1}^2+P_{2}^2)^{1/3}}\right]~,
\end{equation}
where $\dot{\phi}_1 =P_{1}$ and $\dot{\phi}_2=P_{2}$ are both constants. With these conditions, we get, 
\begin{equation} \phi_{0} = -\left(\frac{2}{3\kappa}\right)\left(P_{1}^2+P_{2}^2\right)t^3-Qt\,,
\end{equation}
where $Q$ is a constant as before. A similar calculation holds for the case with potential.  This solution is essentially non-separable in the variables $\{ x,\,y,\, t\}$ 
\begin{figure}[t]
\centering
\includegraphics[width=10cm]{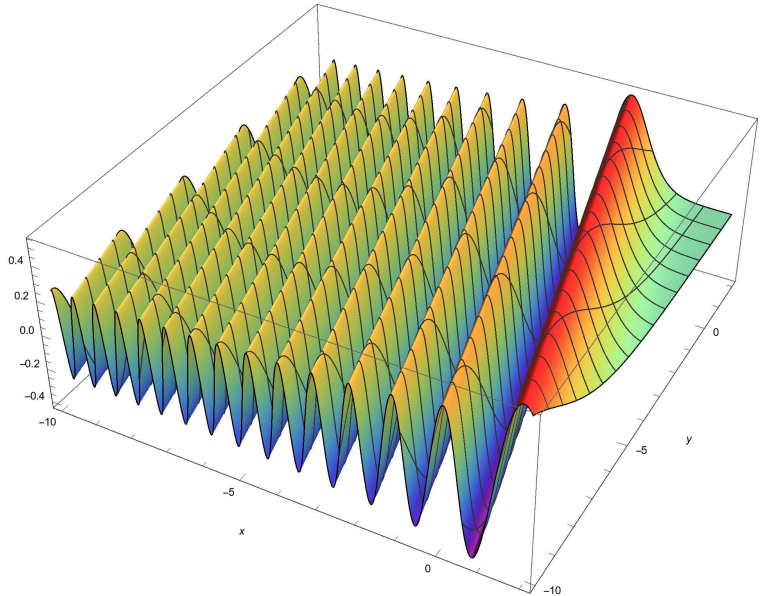}
\caption{The plot of the amplitude in the two dimensional as function of  $x$ and  $y$. Here, we take  $P_1 = 10.1$, $P_2 =2.3 $, $Q = 0$, $\kappa = 1$ and $t = 0$.}
\end{figure}
\section{A nonlinear example}
In this section, we apply this method in solving a particular example of the nonlinear type - the Gross-Pitaevskii equation \cite{pitaevskii}, which is given (in atomic units $m=\hbar=1$) by 
\begin{equation}
 i\psi_t = -\psi_{xx} + V\psi + |\psi|^2\psi ~.
\end{equation}
Here, $V=V(x,t)$ is the potential as given in section-\ref{sec:linpot}. With the substitution of the same trial solution as before, we find that the resulting equations obtained by separating the real and imaginary parts are
\begin{eqnarray}
  A_{xx} - A(\phi_x)^2 - A \phi_t -V A -A^3 &= &0\, \nonumber\\
  A \phi_{xx} + 2 A_x \phi_x + A_t = 0\,.
\end{eqnarray}
The second equation is exactly the same as before given by Eq.~(\ref{eq:Aphi}),  while we get an additional nonlinear term $A^3$ is the first.  With the condition that $\partial^2\phi/\partial x^2$ be zero,
we get, after simplification, the following differential equation
\begin{equation}
 A_{zz} = Az + \mu A^3\,,
\end{equation}
with $z$ defined similarly to the variable $y$ in section 2 (multiplied by an approriate constant) and $\mu$ is a constant. This equation is a variant of the Type-II Painlev\'{e} transcendent \cite{dlmf}, which are known to have solutions in terms of
integral equations involving Airy functions \cite{abl}. Thus we see a specific example of the utility of the present method in obtaining special solutions of nonlinear Schr\"{o}dinger equations.
\section{Discussion}
Here,  we have showed that it is possible to  solve for the nondispersive  solution given by Berry and Balazs in a systematic way. It is possible to obtain the solution  from a purely analytical calculation, without any direct quantum-mechanical arguments.
The method we have presented, which is due to the fact that the second derivative of the phase in the spatial coordinate can be set to zero, leads to an analogy with the modification of the wavefunction
under a Galilean coordinate transformation. This requirement is shown to work in a broad sense, through the derivation of the Airy wavepacket in two dimensions and the 
reduction of the nonlinear Gross-Pitaevskii equation to an ordinary diffenetial equation related to the second Painlev\'{e} transcendent. The latter has analytical solutions and numerical approximations which have a 
striking resemblance with the Airy functions \cite{witt}.

\section*{Acknowledgement}
KRN would like to thank Dr Bhavtosh Bansal and Dr  Vivek Vyas for fruitful discussions. KRN would wish to acknowledge  the Visiting Associateship programme of 
Inter-University Centre for Astronomy and Astrophysics~(IUCAA), Pune.   A part of this work was carried out during the visit to IUCAA under this programme.


\begin{thebibliography}{9}
\bibitem{berry} 
\textit{Nonspreading wave packets}, 
Berry, M. V. and Balazs, N. L., American Journal of Physics, 47, 264-267 (1979), DOI: http://dx.doi.org/10.1119/1.11855

\bibitem{unni}
\textit{Uniqueness of the Airy packet in quantum mechanics}, 
Unnikrishnan, K. and Rau, A. R. P., American Journal of Physics, 64, 1034-1035 (1996), DOI: http://dx.doi.org/10.1119/1.18322

\bibitem{ref:Siviloglou1}
Siviloglou G. A.  and  Christodoulides D. N. , Opt. Lett. 32,
979 (2007).

\bibitem{ref:Siviloglou}
Siviloglou G. A. ,  Broky J. ,  Dogariu A., and  Christodoulides D. N. , PRL  Vol. 99, 213901(2007) 


\bibitem{greenberger2}
\textit{Some remarks on the extended Galilean transformation}, Greenberger, Daniel M., American Journal of Physics, 47, 35-38 (1979), DOI: http://dx.doi.org/10.1119/1.11660

\bibitem{greenberger} 
\textit{Comment on €™Non-spreading wave packets€™}, 
Greenberger, Daniel M., American Journal of Physics, 48, 256-256 (1980), DOI: http://dx.doi.org/10.1119/1.12308


\bibitem{pitaevskii}
L. Pitaevskii and S. Stringari, \textit{Bose-Einstein Condensation and Superfluidity}, Oxford University Press, Oxford, 2016

\bibitem{dlmf}
\textit{NIST Handbook of Mathematical Functions},
Olver, F. W. J., Lozier, D. W., Boisvert, R. F. and Clark, C. W., Cambridge University Press, New York, NY, 2010

\bibitem{abl}
\textit{Exact Linearization of a Painlev\'e Transcendent},
Ablowitz, Mark J. and Segur, Harvey, Phys. Rev. Lett., Vol. 38, Iss. 20, 1103-06, 1977, DOI: http://dx.doi.org/10.1103/PhysRevLett.38.1103

\bibitem{witt}
D Witthaut and H J Korsch 2006 J. Phys. A: Math. Gen. 39 14687 (2006), DOI: http://dx.doi.org/10.1088/0305-4470/39/47/012
\end{thebibliography}
\end{document}